\documentstyle{espart}
\begin{document}
\begin{frontmatter}

\title{Associate Sneutrino-Neutralino/Chargino Production at LEP$\otimes$LHC}

\author{T. W\"ohrmann},
\author{H. Fraas}
\address{Institut f\"ur Theoretische Physik,Universit\"at W\"urzburg,
Am Hubland,\\ 97074 W\"urzburg, Germany}

\begin{abstract}
We examine for representative gaugino-higgsino mixing scenarios
sneutrino-neu\-tral\-ino and sneutrino-chargino production in deep inelastic
ep-scattering at $\sqrt{s}=1.8$ TeV. The cross sections for
sneutrino-chargino production are more than one order of magnitude
bigger than those for sneutrino-squark production. Also for zino-like
neutralinos we find cross sections at least comparable to those for
$\tilde \nu \tilde q$-production.
\end{abstract}
\end{frontmatter}


\section{ Introduction }
Associate production of sleptons and squarks $(ep\rightarrow \tilde l
\tilde q X )$ at HERA and LEP$\otimes $LHC has been extensively
discussed by numerous autors, see e.g. \cite{jls,koma}. If, however, squarks
are heavy but sleptons are
relatively light then these processes are suppressed or even
unaccessible and the associate production of a slepton and a neutralino
$\tilde \chi_{i}^{0} $ or a chargino $\tilde \chi_{i}^{\pm} $ become the
most important SUSY processes at ep-colliders 

{}From the five possible processes which on the parton level are
(i) $ eq \rightarrow \tilde e \tilde \chi_{i}^{0} q $,
(ii) $ eq \rightarrow \tilde e \tilde \chi_{i}^{+} q $,
(iii) $ eq \rightarrow \tilde e \tilde \chi_{i}^{-} q $,
(iv) $ eq \rightarrow \tilde \nu \tilde \chi_{i}^{0} q $,
(v) $ eq \rightarrow \tilde \nu \tilde \chi_{i}^{-} q $, only the
associate production of the lightest neutralino
$\tilde \chi_{1}^{0}$ has
been investigated in the deep inelastic region \cite{ammr},
assuming however, that it is a pure
photino. We have examined for LEP$\otimes$LHC and some
representative SUSY-scenarios the five channels (i) to (v)
and give in this paper some numerical results for associate
sneutrino-neutralino/chargino production.

Distinguished by remarkably large total cross sections and interesting
signatures, which will be -- as well as the standard model background --
briefly discussed, these reactions might be suitable for providing us with a
detectable SUSY signal at LEP$\otimes$LHC. The detailled
discussion of the signatures as well as that of the competing background has
been postponed to a subsequent paper.
\section{Cross Sections and Scenarios}
The Feynman graphs for the basic subprocesses of the reactions
\begin{center}
$ep\rightarrow \tilde
\nu \tilde \chi_{i}^{0} X\, (i=1,\ldots ,4)$,
\end{center}
\begin{center}
$ep\rightarrow \tilde
\nu \tilde \chi_{i}^{-} X\, (i=1,2)$ \end{center}
are shown in fig. 1. \marginpar{{\sf Fig. 1}} The relevant couplings can
be deduced from the interaction Lagrangian of the minimal supersymmetric
extension of the standard model (MSSM), see e.g. \cite{HaKa}. For the
SUSY parameters we assumed as usual $M'/M=\frac{3}{5}\tan \theta_W$,
$m_{\tilde g}=M\sin^2 \theta_W \alpha_s /\alpha_{em}\simeq 3M$,
with $\sin^2 \theta_W=0.228$ and $\alpha_s=0.1$, and
$\tan \beta=v_2 / v_1 =2$ (the numerical
results are not very sensitive to the value of $\tan \beta$).
For the masses of the gauge bosons we have used $m_Z=91.2$ GeV,
$m_W=80.1$ GeV.

We shall present
numerical results at $\sqrt{s}=1.8$ TeV for two different mixing
scenarios shown in table \ref{tab}.\marginpar{{\sf Table 1}}
For each of these mixing scenarios cross sections
have been calculated for two different relations between sneutrino mass and
squark mass: $m_{\tilde q}=m_{\tilde \nu}$ in scenarios (A.1) and (B.1)and
$m_{\tilde q}=4\cdot m_{\tilde \nu}$ in scenarios (A.2) and (B.2).
The last two scenarios (A') and (B') with $m_{\tilde
\nu}=m_{\tilde g}$ and $m_{\tilde q}=\sqrt{2}\cdot m_{\tilde \nu}$ are
motivated by renormalization group relations coupling the sfermion
masses and the gaugino mass parameter $M$ of the MSSM \cite{Pol}.
Allowing an error
of at most 4\% for the sneutrino and squark masses this choice is for
values of $M$ between 45 GeV and 450 GeV (and $\tan \beta =2$)
compatible with the mass relations given in \cite{Pol}. The value of
$\mu$ in scenarios (A') and (B') is the same as in scenarios (A)
and (B), respectively. Notice, however, that in these scenarios both
the mass and the mixing character of the neutralinos and charginos
depend on the respective value of the sneutrino mass.

For the momentum transfer square  $Q^2=-(p_{q_{out}}-p_{q_{in}})^2$
to the quark we impose a cut with $Q^{2}_{cut}= 10(\mbox{GeV})^2$.
While the dependence on $Q^{2}_{cut}$ is approximately logarithmic for graphs
(a), (b), (e) with photon exchange those with exchange of massive gauge bosons
and charginos are not very sensitive on this cut.
The final step in the evaluation of the cross sections consists in
folding those for the parton subprocesses with the quark
distribution functions $f_q (x,\tilde Q^2)$ of Gl\"uck, Reya and Vogt
\cite{GRV} for the valence quarks and the {\em u, d, s} sea quarks.
The situation is somewhat more complicated than for slepton-squark
production since in our case the momentum transfer to the nucleon
depends on the respective reaction mechanism in fig. 1. We have,
however, numerically checked that in the kinematic region investigated
here $\tilde Q^2=\frac{1}{2}(sx-(m_{\tilde \nu }+m_{\tilde \chi})^{2})$
is a satisfactory approximation. The errors involved are less than
10\%.

For the squark width entering into the graph (c) all contributions from
two-body decays have been taken into account.
The integration was performed with the Monte-Carlo program {\em vegan}.
\section{Numerical Results}
\subsection{The process $ep\rightarrow \tilde \nu \tilde
\chi_{i}^{0} X$ }
In figs. 2--5 we show the total production cross sections $\sigma
(ep\rightarrow \tilde \nu \tilde \chi_{i}^{0} X ),\,
i=1,\ldots,4,$ as a function
of the sneutrino mass for the scenarios (A.1), (A.2), (B.1) and (B.2)
at $\sqrt{s}=1.8$ TeV. For comparison we give in the figs.
also the corresponding cross section
$\sigma (ep\rightarrow \tilde \nu \tilde q X )$ for associate
sneutrino-squark production. We give no figs. for scenarios (A') and (B')
since only for (A')
the cross sections are only for the LSP $\tilde \chi^{0}_{1} $ and
$m_{\tilde \nu} \leq 350$ GeV bigger than $10^{-2}$
pb (between $10^{-2}$ and 0.1 pb).
The cross sections for scenario (B') are smaller than $10^{-2}$pb.

As a consequence of the large $W$-couplings in the Feynman graphs (a),
(b), (e) of fig. 1 the cross sections for
zino like neutralinos are comparable to (scenario (B))
or even considerably higher (scenario (A)) than those for
$\tilde \nu \tilde q$-production. Therefore the question which of the
four neutralinos will be produced with the highest rate sensitively
depends on the mixing scenario. Thus in scenario (A.2)
the cross section for $\tilde \chi_{2}^{0}$ and in scenario (B.2)
even that for the heaviest neutralino is the dominating one
being ten times as large as that for the lightest neutralino
$\tilde \chi_{1}^{0}$.

Since for pure photinos graph (b) does not contribute, the cross section
for photino like neutralinos is much smaller than that for zino like
neutralinos. For larger selectron masses also graph (a) is suppressed by
the selectron propagator. The steep ascent of the cross sections for
$\tilde \chi_{1}^{0}$ in scenario (A.1) and $\tilde \chi_{3}^{0}$ in
scenario (B.1) originates from the
contribution of graph (c), where for $m_{\tilde q}=m_{\tilde \nu}$ the squark
approaches its mass shell in the accessible region of the phase space.
Even the relatively small gaugino components of the
light neutralinos
$\tilde \chi_{1,2}^{0}$ in (B.1) suffice to produce this step in the
cross sections. If, however, the zino component is strong enough, this
effect will be suppressed by the strong contributions of the Feynman
diagrams (a), (b), (e).

Since for pure higgsinos only graph (e) contributes, the cross sections for
the heavy higgsino-like neutralinos $\tilde \chi_{3,4}^{0}$
in scenarios (A.1) and (A.2) are nearly independent of the squark mass.
\marginpar{{\sf Figs. 2--5}}
\subsection{The process $ep \rightarrow \tilde \nu \tilde \chi^{-}_{i}
X$}
The total cross sections $\sigma(ep \rightarrow \tilde \nu \tilde
\chi^{-}_{i} X),\, i=1,2$ are shown in figs. 6--8. Due to the
dominance of the
contributions from gauge boson exchange in graphs (a), (b), (e),
the numerical results are nearly
identical for $ m_{\tilde q}=m_{\tilde \nu}$ and $ m_{\tilde
q}=4\cdot m_{\tilde \nu}$. We therefore give the results for scenarios
(A.1),(B.1) and (B') only. We give no results for scenario (A') since
these results are quite similar to those in scenario (B').
Contrary to $\tilde \nu \tilde q
$-production all partons are contributing to
$\tilde \nu \tilde \chi_{i}^{-} $-production. Together with the strong
$Z$-couplings in graphs (a), (b), (e) and photon couplings in graphs
(b), (e) this leads for all scenarios to cross sections for $\tilde \nu
\tilde \chi_{i}^{-} $-production being between one and two orders of
magnitude larger than those for $\tilde \nu \tilde q $-production.
The cross section is the highest for the light wino-like $\tilde
\chi_{1}^{-}$ in scenario (A.1) but even
for the heavy wino-like $\tilde \chi_{2}^{-}$ in (B.1) it is considerably
larger than that for
$\tilde \nu \tilde q $-production and also larger than that for the
light higgsino-like state $\tilde \chi_{1}^{-}$, with substantial
contributions from graph (e) only.

In contrast to sneutrino-neutralino production also for scenarios
(A') and (B') the cross sections for $\tilde \nu \tilde \chi_{i}^{-}
$-production are larger than $10^{-2}$ pb in a wide range of
parameter space
and considerably larger than those for $\tilde \nu \tilde q
$-production. Notice that in scenarios (A') and (B') the
mass values as well as the couplings of both chargino states are varying
with increasing sneutrino mass: For the lower values of $m_{\tilde \nu}$
the light chargino is wino-like and the heavy chargino is
more higgsino-like. The situation changes with increasing $m_{\tilde \nu}$
so that the light chargino becomes more and more higgsino-like whereas
the heavy one becomes more and more wino-like. Simultaneously the mass of
the heavy chargino is rapidly increasing, whereas that of the light
chargino asymptotically approaches the value $|\mu |$. This
interplay between mass and mixing character produces the two crossings of the
cross sections in fig. 8 for scenario (B').\marginpar{{\sf
Figs. 6--8}}
\section{ Signatures }
In order to study the possible signals for associate
sneutrino-neutralino/chargino production it is indispensable to include
the decay of these particles as well as a discussion of the competing
standard model background. Here we shall restrict ourselves to some
remarks, postponing a more detailled discussion of signatures and
background to a subsequent paper.

Light supersymmetric particles decay directly into the lightest neutralino
(which is assumed to be the lightest supersymmetric particle LSP and
stable) and fermions, whereas heavy sparticles decay over complex cascades
ending at the LSP. These cascade decays of heavy sparticles
have two important consequences. On the one hand they will lead to
events with besides one or several leptons, jets and missing energy one
or two $W$ or $Z$ bosons in the final state \cite{tata}. On the other
hand they can significantly enhance the possible signals of the
respective process \cite{cuyp}. The actual decay patterns and the
dominant signatures will, however, sensitively depend on the
supersymmetric parameters and the slepton mass. Thus in scenarios (A.1)
and (A.2) where the cross sections are the biggest for the processes
$ep\rightarrow \tilde \nu \tilde \chi_{2}^{0} X $ and
$ep\rightarrow \tilde \nu \tilde \chi_{1}^{-} X $
the dominant signatures are $ej\not \! E$, $2ej\not \! E$ and
$3ej\not \! E$ for $m_{\tilde l}=100$ GeV and $ej\not \! E$,
$2ej\not \! E$ for $m_{\tilde l}=500$ GeV.
For scenarios (B.1) and (B.4) on the other hand with
$ep\rightarrow \tilde \nu \tilde \chi_{4}^{0} X $ and
$ep\rightarrow \tilde \nu \tilde \chi_{2}^{-} X $ as the dominant
processes the favored signatures are $ej\not \! E$ and $Wj\not \! E$ for
$m_{\tilde l}=100$ GeV and $eWj\not \! E$ and $e2Wj\not \! E$
for $m_{\tilde l}=500$ GeV ($j$ denotes an arbitrary number of jets).

The most important sources of background are single $W$ and $Z$
production $ep\rightarrow \nu W X,\nu ZX$ and $ep\rightarrow eW X,eZX$
followed by the decays $W\rightarrow l\nu_l $, $Z\rightarrow \nu_l \bar \nu_l
$ and $Z\rightarrow l^+l^-
$ giving rise to events with one, two or three charged
leptons \cite{ammr,baur}. On the other hand the case of single top
production $ep \rightarrow \nu \bar t b X$ followed by the decay $\bar t
\rightarrow \bar b W^-$ gives rise to the $Wj\not \! E$ configuration and the
neutral current process $ep\rightarrow et\bar t X$ is a source of the
background for $e2Wj\not \! E$ events \cite{ali}. Since, however, the cross
section for $t\bar t$-production is rather small ($\simeq 0.06$ pb for
$\sqrt{s}=1260$ GeV and $m_t =180$ GeV), one would expect that this is
the least dangerous of the competing standard model backgrounds.
Detailled Monte Carlo studies taking into account the background are
needed to assess the observability of the SUSY signal from associate
sneutrino-neutralino/chargino production.
\section{ Conclusion}
We have computed for representative gaugino-higgsino mixing scenarios
the total cross sections for associate sneutrino-neutralino and
sneutrino-chargino production at LEP$\otimes $LHC. For wino-like
charginos $\tilde \chi_{i}^{-},\, i=1,2$ as well as for higgsino-gaugino
mixtures they are between one and two
orders of magnitude bigger than those for $\tilde \nu \tilde q
$-production: about 0.1pb for $m_{\tilde \nu}=500$GeV and between 1pb
and 10pb for $m_{\tilde \nu}=50$GeV. The cross sections for a light
higgsino-like chargino are still one order of magnitude higher than those
for $\tilde \nu \tilde q $-production.

The situation is less favorable for sneutrino-neutralino production.
Here the cross sections are the highest for neutralinos with a strong
zino component. For $m_{\tilde q}=m_{\tilde \nu}$ the cross sections for
a zino-like neutralino are generally comparable to those for $\tilde \nu
\tilde q $-production. If, however, the squarks are appreciably heavier
than the sneutrinos ($m_{\tilde q}=4\cdot m_{\tilde \nu}$) then in a mass
region where the $\tilde \nu \tilde q $-cross section has already
decreased to $10^{-4}$pb that for a zino-like neutralino is still
between $10^{-2}$pb and 0.1pb.

Similar as for chargino production the question which of the neutralino
cross sections is the dominating one depends much more on the mixing
properties than on the mass of respective states. Thus in our scenario
(B.2) even the cross section for the heaviest neutralino is for
$m_{\tilde \nu}\geq 150$GeV the dominating one, being one order of
magnitude beyond that for the lightest neutralino and surmounting that
for $\tilde \nu \tilde q $-production.

Together with the subsequent decays of the produced sparticles this
leads to interesting signatures consisting in up to four charged
leptons, hadronic jets, missing energy and in case of scenario (B)
massive gauge bosons. A quantitative analysis of these signatures will be
postponed to a subsequent paper.

The size of the cross section obtained for sneutrino-chargino production,
comparable to or even bigger than that for competing standard model processes,
let us however suggest, that this process should provide an attractive
channel in the search for supersymmetric events at LEP$\otimes $LHC.
\section*{ Acknowledgements }
The authors would like to thank F.~Franke for many helpful discussions.
T.W. was supported by the {\em Friedrich-Ebert-Stiftung}. All numerical
calculations were performed at the
{\em Rechenzentrum der Universit\"at W\"urzburg}.
 \newpage
\section*{Table Captions}
\begin{table}[h]
\caption{\label{tab}
Neutralino and chargino states for scenarios (A.1), (A.2) and
(B.1), (B.2) in terms of the weak eigenstates ($\tilde
\gamma, \tilde Z, \tilde H^{0}_{a},\tilde H^{0}_{b}$) for neutralinos
and ($-i\lambda^+, \psi^{1}_{H_2}$) for charginos (see
\protect\cite{bartl}  for details of gaugino-higgsino mixing).}
\end{table}
\section*{Figure Captions}
Fig. 1: Feynman diagrams for the basic subprocesses $eq
\rightarrow \tilde \nu \tilde \chi_{i}^{0} q' \, (eq
\rightarrow \tilde \nu \tilde \chi_{i}^{-} q)$
\newline
Fig. 2: Cross sections in scenario (A.1) for the processes $ep
\rightarrow \tilde \nu \tilde \chi_{i}^{0} X$, with dashed line for
$i$=1, dotted line for $i$=2, dash-dotted for $i$=3, dash-dot-dot for
$i$=4 and solid line for $ep
\rightarrow \tilde \nu \tilde q X$.\newline
Fig. 3: The same as fig. 2 for scenario (A.2).\newline
Fig. 4: The same as fig. 2 for scenario (B.1).\newline
Fig. 5: The same as fig. 2 for scenario (B.2).\newline
Fig. 6: Cross sections in scenario (A.1) for the processes $ep
\rightarrow \tilde \nu \tilde \chi_{i}^{-} X$, with dashed line for
$i$=1, dotted line for $i$=2 and solid line for $ep
\rightarrow \tilde \nu \tilde q X$.\newline
Fig. 7: The same as fig. 6 for scenario (B.1).\newline
Fig. 8: The same as fig. 6 for scenario (B'), with the
masses of the charginos $\tilde \chi_{1}^{-}$ and $\tilde
\chi_{2}^{-}$ included.\newpage
\begin{tabular}{|c||c|c|}
\hline
 & A  & B \\ \hline \hline
$\tan \beta$ & 2 & 2 \\ \hline
$\mu$ & --219 GeV &  --44 GeV \\ \hline
M & 73 GeV &  219 GeV \\ \hline
$\tilde \chi_{1}^{0}$ & $m=40$ GeV& $m=40$ GeV\\
 & (--0.95,+0.30,+0.08,+0.08) & (--0.06,+0.13,--0.18,+0.97) \\ \hline
$\tilde \chi_{2}^{0}$ & $m=88$ GeV& $m=74$ GeV\\
 & (--0.32,--0.89,--0.18,--0.27) & (+0.07,--0.33,+0.92,+0.22) \\ \hline
$\tilde \chi_{3}^{0}$ & $m=225$ GeV& $m=118$ GeV\\
 & (+0.02,+0.20,+0.35,--0.92) & (+0.92,--0.32,--0.20,+0.06) \\ \hline
$\tilde \chi_{4}^{0}$ & $m=244$ GeV& $m=243$ GeV\\
 & (+0.01,--0.27,+0.92,+0.29) & (+0.37,+0.88,+0.29,--0.04) \\ \hline
$\tilde \chi_{1}^{+}$ & $m=87$ GeV&  $m=61$ GeV\\
 & (+0.99,+0.10) &  (+0.39,--0.92) \\ \hline
$\tilde \chi_{2}^{+}$ & $m=241$ GeV&  $m=242$ GeV\\
 & (--0.10,+0.99) &  (+0.92,+0.39) \\ \hline
\end{tabular}
\newline
\begin{center}
Table 1
\end{center}
\end{document}